\documentclass[prb,twocolumn,floatfix,showpacs,amssymb]{revtex4}
\usepackage{graphicx}
\usepackage{dcolumn}
\usepackage{bm}

\begin{document}

\title{Microscopic theory of equilibrium properties of\\
F/S/F trilayers with weak ferromagnets}

\author{R. M\'elin\footnote{E-mail: melin@grenoble.cnrs.fr}}
\affiliation{ 
Centre de Recherches sur les Tr\`es Basses
Temp\'eratures (CRTBT)\footnote{U.P.R. 5001 du CNRS, Laboratoire conventionn\'e
avec l'Universit\'e Joseph Fourier},\\ Bo\^{\i}te Postale 166, F-38042 Grenoble Cedex 9,
France}

\begin{abstract}
The aim of this paper is to explain the non monotonic temperature dependence of the
self-consistent superconducting gap of ferromagnet/superconductor/ferromagnet (F/S/F)
trilayers
with weak ferromagnets in the parallel alignment (equivalent to F/S bilayers).
We show that this is due to 
Andreev bound states that
compete with the formation of a minigap.
Using a recursive algorithm
we discuss in detail the roles of various parameters
(thicknesses of the superconductor and ferromagnets,
relative spin
orientation of the ferromagnets, exchange field,
temperature,
disorder, interface transparencies). 
\end{abstract}


\maketitle

\newcommand{\Ht}{\hat{t}}
\newcommand{\Hg}{\hat{g}}
\newcommand{\HG}{\hat{G}}
\newcommand{\HI}{\hat{I}}
\newcommand{\Hh}{\hat{h}}
\newcommand{\be}{\begin{equation}}
\newcommand{\ee}{\end{equation}}
\newcommand{\HK}{\hat{K}}
\newcommand{\HSigma}{\hat{\Sigma}}
\newcommand{\nonb}{\nonumber}
\newcommand{\Hu}{\hat{u}}

\section{Introduction}

In conventional
superconductivity the attractive interaction mediated by phonons
binds electrons into Cooper pairs that condense in
the BCS ground state with a zero temperature gap
$\Delta$ to the first quasiparticle
excitations~\cite{Tinkham}. In ferromagnetism
electron interactions generate a spin symmetry breaking that can
be described by the Stoner model in which electrons subject
to an exchange field $h_{\rm ex}$ acquire a Zeeman energy.

Many physical phenomena are involved at the interfaces between
superconductors (Ss) and ferromagnets (Fs). For instance it was
shown in the early 1970's
that the Fermi surface spin polarization of a ferromagnetic
metal could be measured by spin-resolved tunneling between
a ferromagnet and a superconducting film in the presence
of Zeeman splitting\cite{Tedrow}.
The Fermi surface spin polarization was 
measured more recently\cite{Soulen,Upadhyay}
by Andreev reflection at F/S interfaces\cite{deJong-Beenakker}
with highly transparent interfaces, not in thin film geometries.
The non equilibrium spin population in the ferromagnet 
plays also a role in Andreev reflection at
F/S interfaces\cite{Falko,Jedema,Belzig}.
Andreev reflection with Zeeman splitting
in a thin film geometry could be another way
to probe the Fermi surface spin polarization\cite{Melin-Europhys}.

These experiments can be interpreted without discussing the
proximity effect (the pair correlations induced in the
normal metal or ferromagnet)
that has focused an
important interest recently.
It is well established both theoretically and experimentally that
the pair amplitude induced in
a ferromagnetic metal oscillates in space and can become
negative\cite{Fulde-Ferrel,Larkin,Clogston,Demler,Buzdin1},
giving rise to the $\pi$-coupling.
The $\pi$-coupling generates 
oscillations of the critical temperature of
F/S multilayers as a function of the
thickness of the ferromagnetic
layers\cite{Buzdin2,Radovic,exp1,exp2,exp3}. 
The bad quality of interfaces may however
also play a role in these experiments\cite{transpa,Zareyan}.
Direct evidences of the $\pi$-coupling
have been obtained recently\cite{Ryazanov,Kontos,Gandit,Sellier}.
The proximity effect at F/S interfaces is related to the
formation of the so-called
Andreev bound states. Andreev bound states were first discussed
by de Gennes and Saint-James\cite{deGennes-bound} 
and Andreev\cite{Andreev}
for normal metal (N)/S interfaces.
Several theoretical investigations of Andreev bound states
at F/S interfaces
have been presented recently\cite{Vecino,Zareyan,Cserti}
as well as numerical investigations of the proximity effect at F/S
interfaces, based on simulations of the Bogoliubov-de Gennes
equations\cite{Valls}.

The interest in the inverse
proximity effect (the properties of the superconductor)
at F/S interfaces dates back to the
1960's\cite{deGennes-FSF,Deutscher-Meunier,Hauser} where it was shown that
with insulating ferromagnets an exchange field
is induced in the superconducting electrode of a F/S/F trilayer
in the parallel alignment.
The theoretical prediction by de Gennes\cite{deGennes-FSF} was
further confirmed by experiments in the late 1960's
with metallic\cite{Deutscher-Meunier}
and insulating\cite{Hauser} ferromagnets.
Recent experiments were carried out with metallic ferromagnets,
which confirmed the effect on the critical temperature\cite{Gu}.
It was shown recently\cite{Jirari,Volkov}
that an exchange field
in the superconductor
exists also with metallic
ferromagnets but the sign of the magnetization in the
superconductor is opposite
to the magnetization in the ferromagnet.
An exchange field in a superconductor
is pair breaking.
As a consequence
the superconducting transition temperature in the
parallel alignment is smaller than in the antiparallel
alignment.

With metallic ferromagnets the proximity effect
may influence the value of the superconducting gap and
transition temperature.
It was realized
recently\cite{Melin-FSF,Jirari,Baladie,Buzdin-Daumens,Melin-Feinberg-FSF}
that the
proximity effect in F/S/F trilayers is quite special since
under some conditions the
zero temperature superconducting gap in the parallel alignment 
can be
larger than in the antiparallel alignment. This was
established\cite{Melin-FSF,Jirari} within a model of multiterminal hybrid
structure originally proposed for transport
properties\cite{Deutscher-Feinberg,Falci,Melin-Feinberg-tr,Melin-Peysson,Lambert}
and is related to spatially separated
superconducting correlations among the two ferromagnets.
The same behavior was found 
within a model of F/S/F trilayer with atomic thickness
and half-metal ferromagnets\cite{Buzdin-Daumens},
which was finally
extended to Stoner ferromagnets\cite{Melin-Feinberg-FSF}.

Other physical effects take place with weak ferromagnets
for which the exchange field is smaller than the superconducting
gap, as indicated by the reentrance of
the critical temperature of F/S superlattices as a function of the
exchange field\cite{Andreev-Buzdin,Houzet}, and by the reentrance of
the critical temperature of F/S bilayers as a function of the thickness
of the ferromagnet\cite{Baladie,Fominov,Khusainov}.
The full temperature dependence of the self-consistent superconducting gap
was calculated in Ref.~\onlinecite{Melin-Feinberg-FSF} and it was shown that 
for F/S bilayers or F/S/F trilayers with atomic thickness 
the superconducting gap can have a non monotonic temperature dependence:
within a given range of parameters
the superconducting gap first increases as temperature $T$ is reduced,
reaches a maximum, and decreases to zero as $T$ is further reduced. 
Within a narrow range of interface transparencies there exists also
a reentrant behavior
of the superconducting gap at low temperature\cite{Melin-Feinberg-FSF}.

The purpose of this article is to show that this behavior is related to the
formation of Andreev bound states that can compete with the formation of
a superconducting minigap. We find a systematic correlation between
Andreev bound states at the Fermi level and a reduction of the
low temperature self-consistent superconducting gap, which
constitutes the main result of this article. 

The article is organized as follows.
Preliminaries are given in section~\ref{sec:prelim} in which
we discuss the models and the Green's functions formalism.
A recursive algorithm, more efficient that the direct
inversion of the Dyson matrix\cite{Melin-Feinberg-FSF}, is
presented in section~\ref{sec:recur}. 
Different regimes can be obtained depending on how the lateral
dimensions $L_S$ of the superconductor and $L_F$ of the ferromagnets
compare respectively (i)
to the zero temperature lateral superconducting coherence length
$\xi_S^{(\perp)}=2 t_0 a_0/ \pi \Delta$\cite{Vecino}, where $t_0$ is the
lateral hoping amplitude, and $a_0$ is equal to the interatomic distance
in the tranverse direction; and (ii) to the zero temperature 
lateral exchange length
$\xi_F^{(\perp)}=2 t_0 a_0/ \pi h_{\rm ex}$\cite{Vecino}. 
We present a detailed investigation of the different
regimes $L_F,L_S \ll \xi_S^{(\perp)}$ (section~\ref{sec:small});
$L_S \alt \xi_S^{(\perp)}$ and $L_F \ll \xi_F^{(\perp)}$ (section~\ref{sec:finite});
$L_S \ll \xi_S^{(\perp)}$ and $L_F \agt \xi_F^{(\perp)}$ (section~\ref{sec:corre}).
We cannot carry out a systematic study of the regime
$L_S \sim \xi_S^{(\perp)}$ , $L_F \agt \xi_F^{(\perp)}$ which is too
demanding from a computational point of view.
In section~\ref{sec:compa} we provide a comparison with
other models proposed recently to describe Andreev bound states in
F/S hybrids\cite{Cserti,Vecino}.
Concluding remarks are given in section~\ref{sec:conclu}.

\section{Preliminaries}
\label{sec:prelim}
\subsection{The models}
\begin{figure}
\includegraphics [width=.7 \linewidth]{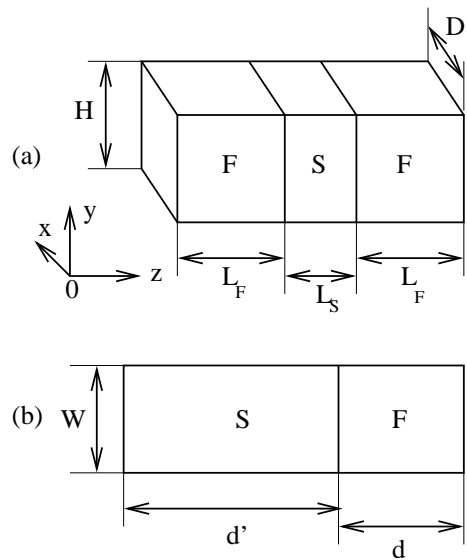}
\caption{(a) Schematic representation of the F/S/F trilayer. 
$D$ and $H$ are sent to to infinity 
(infinite planar geometry).
The ferromagnets
(superconductor) are made of $L_F$ ($L_S$)
layers stacked in the lateral ($z$) direction.
(b) Schematic representation of the 2D F/S interface
model\cite{Cserti} considered
in section~\ref{sec:2D}. $d'$ is sent to infinity.
\label{fig:schema1_FSF}
}
\end{figure}
\begin{figure}
\includegraphics [width=.9 \linewidth]{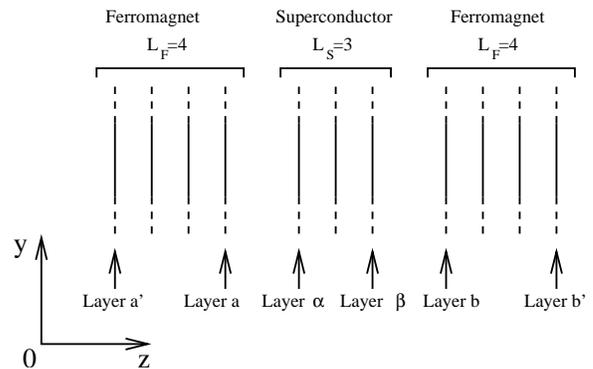}
\caption{Cut in the (z,y) plane of the F/S/F trilayer 
with $(L_S/a_0,L_F/a_0)=(3,4)$, where $a_0$ is the
interplane spacing.
The ferromagnets and the superconductor
are infinite in the $x$ and $y$ directions.
The left ferromagnet ends at the layers $a'$ and $a$.
The right ferromagnet ends at the layers $b$ and $b'$.
The superconductor ends at layers $\alpha$ and $\beta$.
\label{fig:schema2_FSF}
}
\end{figure}

We suppose that the two ferromagnets of the F/S/F trilayer have the same
thickness $L_F$.
We note $L_S$ the thickness of the
superconductor (see Fig.~\ref{fig:schema1_FSF}).
The three electrodes are made of 2D planes stacked along the $z$ axis
(see Fig.~\ref{fig:schema2_FSF}).
Each superconducting plane is described by the BCS Hamiltonian
\begin{eqnarray}
\label{eq:H-BCS}
{\cal H}_{\rm BCS}^{(2D)} &=& \sum_{ {\bf k},\sigma}
\epsilon(k) c_{ {\bf k},\sigma}^+ c_{{\bf k},\sigma}\\
&+& \Delta \sum_{\bf k} \left( 
c_{{\bf k},\uparrow}^+ c_{-{\bf k} , \downarrow}^+ +
c_{{\bf k} , \downarrow} c_{-{\bf k},\uparrow} \right)
\nonb
,
\end{eqnarray}
where the wave vector ${\bf k}$ is parallel to the 2D layer.
The free electron dispersion relation within one layer is 
$\epsilon(k)=\hbar^2 k^2 / 2 m$. 
The ferromagnets
are described by the Stoner model
\begin{eqnarray}
\label{eq:H-Stoner}
{\cal H}_{\rm Stoner}^{(2D)} &=& \sum_{ {\bf k} ,\sigma}
\epsilon(k) c_{ {\bf k},\sigma}^+ c_{{\bf k},\sigma}\\
\nonb
&-&h_{\rm ex} \sum_{\bf k} \left(
c_{{\bf k},\uparrow}^+ c_{{\bf k} , \uparrow} -
c_{{\bf k} , \downarrow}^+ c_{{\bf k},\downarrow} 
\right)
,
\end{eqnarray}
where $h_{\rm ex}$ is the exchange field.
The coupling between two planes belonging to the same superconducting
or ferromagnetic electrode is given by
\be
\label{eq:tunnel-t0}
{\cal W}_{\rm F-F} = {\cal W}_{\rm S-S} = t_0
\sum_{ {\bf x}_L,\sigma}  \left(
c_{ {\bf x}_L,\sigma}^+ c_{ {\bf x}_R,\sigma}
+ c_{ {\bf x}_R,\sigma}^+ c_{ {\bf x}_L,\sigma}
\right)
,
\ee
whereas the coupling between the ferromagnets and the superconductor
is given by
\be
\label{eq:tunnel-t}
{\cal W}_{\rm F-S} = t \sum_{ {\bf x}_L,\sigma} \left(
c_{ {\bf x}_L,\sigma}^+ c_{ {\bf x}_R,\sigma}
+ c_{ {\bf x}_R,\sigma}^+ c_{ {\bf x}_L,\sigma}
\right)
,
\ee
where the summation runs over all sites at the interface
(${\bf x}_L$ corresponds to a site on the left side and
${\bf x}_R$ is the corresponding site on the right side).
The parameters are such that the coherence length within one layer
$\xi_S^{(\parallel)}=\hbar v_F^{(\parallel)}/\Delta$ and the tranverse
coherence length $\xi_S^{(\perp)}=2 t_0 a_0/\pi \Delta$ are larger than
the width $L_S$ of the superconductor. 
The model should stricktly speaking apply to ballistic systems whereas
real samples are usually in the diffusive regime. However we
can include disorder in one particular case (the F/S/F trilayer with
atomic thickness, see section~\ref{sec:dis}).
In this case the qualitative behavior
is robust against increasing disorder. Fabry-Perot resonances
generate parity effects for small values of $L_S$ and $L_F$ in the
ballistic model. These parity effects are not expected to occur in
the diffusive regime, but do not occur either in our simulations
if $L_S$ and $L_F$ are sufficiently large. By increasing
$L_F$ and $L_S$ we obtain a cross-over between the regime
$L_F,L_S \alt \lambda_F$ and the regime $\lambda_F \alt L_F, L_S
\ll \xi_S^{(\perp)}$. We find the same qualitative effects in the
two regimes. Moreover
we obtain in section~\ref{sec:finite}
a non monotonic variation of the self-consistent superconducting gap
as a function of the exchange field. This is compatible with the non
monotonic variation of the critical temperature as a function of the
exchange field\cite{Baladie,Fominov,Khusainov} obtained in the
context of linearized Usadel equations for disordered conductors.
The compatibility between the two behaviors indicates the
validity of our approach.

\subsection{Green's functions}

\subsubsection{Zero temperature Green's functions}

The Green's functions of an isolated superconductor can be gathered
in a $4 \times 4$ matrix in the spin $\otimes$ Nambu
representation but in the absence of non collinear
magnetizations\cite{Melin-Peysson,Bergeret-2001,Kadi,Chte}
the quantization axis can be chosen parallel
to the exchange field so that the $4 \times 4$ Green's functions reduce
to two separate $2 \times 2$ matrices, one in each spin sector.
For practical purpose we work in the spin-up sector.
The Green's function is given by
\begin{eqnarray}
\label{eq:Green-def}
&& \hat{g}_{{\bf x},{\bf y}}(t,t') = -i\\
\nonb
&& \left( \begin{array}{cc}
\langle T_t \left( c_{{\bf x},\uparrow}(t) , c_{{\bf y},\uparrow}^+(t') \right) \rangle &
\langle T_t \left( c_{{\bf x},\uparrow}(t) , c_{{\bf y},\downarrow}(t') \right) \rangle \\
\langle T_t \left( c_{{\bf x},\downarrow}^+(t) , c_{{\bf y},\uparrow}^+(t') \right) \rangle &
\langle T_t \left( c_{{\bf x},\downarrow}^+(t) , c_{{\bf y},\downarrow}(t') \right) \rangle
\end{array} \right)
,
\end{eqnarray}
where ${\bf x}$ and ${\bf y}$ are two arbitrary sites in the superconductor
and $T_t$ is the usual $T$-product\cite{Abrikosov}.
The ``11'' component describes the propagation of a spin-up electron,
the ``22'' component describes the propagation of a spin-down hole
and the ``12'' and ``21'' components describe superconducting correlations.
After Fourier transforming 
we obtain a standard expression for the different elements
of the Green's function~\cite{Abrikosov}:
\begin{eqnarray}
\label{eq:galpha11}
g_{\alpha,\alpha}^{1,1}(\xi,\omega)&=&
\frac{u_k^2}{\omega-E_k+i\eta}
+\frac{v_k^2}{\omega+E_k-i\eta}\\
\label{eq:falpha12}
f_{\alpha,\alpha}^{1,2}(\xi,\omega)&=&
-\frac{\Delta}
{\left[\omega-E_k+i\eta \right]
\left[\omega+E_k-i\eta \right]}
.
\end{eqnarray}
The variable $\xi$ is related to the kinetic energy:
$\xi_k =\hbar^2 k^2/(2m)-\epsilon_F$ where
$\epsilon_F=\hbar^2 k_F^2/2m$ is the Fermi energy.
$E_k=\sqrt{\Delta^2+\xi_k^2}$ is the quasiparticle energy
and $u_k^2=(1+\xi_k/E)/2$ and
$v_k^2=(1-\xi_k/E)/2$ are the BCS coherence factors.

The Green's function of a spin-up ferromagnet
is diagonal in Nambu space. The ``11'' component is given by
\be
g_{a,a}^{1,1}(\xi,\omega)
= \frac{1}{[\omega-\xi+h_{\rm ex} +i \eta \mbox{ sgn}(\xi-h_{\rm ex})]}
,
\ee
and the Green's functions of a spin-down ferromagnet are obtained by changing 
$h_{\rm ex}$ into $-h_{\rm ex}$.

The Green's functions $\HG_{{\bf x},{\bf y}}$
of the connected trilayer are given by the
Dyson equation $\HG=\Hg+\Hg\otimes \HSigma \otimes \HG$,
where in a compact notation $\HSigma$ is the self-energy
corresponding to the tunnel Hamiltonian~(\ref{eq:tunnel-t}) and $\otimes$
corresponds to a summation over spatial variables and a convolution
over time variables. We look for non perturbative solutions
of the Dyson equations suitable for describing Andreev bound states.

\subsubsection{Finite temperature Green's functions}

Finite temperature Green's functions are obtained through the 
analytic continuation $\omega \rightarrow i \omega$ and by
summing over the Matsubara frequencies $\omega_n
=(2 n + 1) \pi T$ where $T$ is the temperature\cite{Abrikosov}.
The superconducting gap is determined from the
BCS self-consistency equation\cite{Abrikosov}
\begin{equation}
\label{eq:self}
\Delta_{\bf x}=\lambda T \sum_n \int
\frac{d^2{\bf k}}{(2\pi)^2} G^{1,2}_{{\bf x},{\bf x}}({\bf k},i \omega_n)
,
\end{equation}
where $\lambda$ is the
strength of the attractive electron-electron interaction.
To evaluate (\ref{eq:self}) we change variable to
$\xi=\hbar^2 k^2/(2m)-\epsilon_F$ and restrict the integral to
$|\xi|<\omega_D$. To avoid introducing new parameters we use
$\omega_D=\epsilon_F=\hbar^2 k_F^2/ 2 m$.
We note $\Delta_0$ the superconducting gap of an isolated 2D layer.
All energy scales will be compared to $\Delta_0$.

\section{Recursive algorithm for the F/S/F trilayer}
\label{sec:recur}

\subsection{Green's functions of an isolated electrode}
We aim to describe F/S/F
trilayers with a finite thickness of all electrodes. In this
respect the
$(L_S/a_0,L_F/a_0)=(1,1)$ trilayer is just viewed as a toy-model:
we use a mean field approach that does not incorporate 
the phase fluctuations of the order parameter and we
do not consider possible instabilities such as spin or charge
density wave. 

The ferromagnetic and superconducting electrodes of the F/S/F trilayer
consist of a finite number of layers stacked along the $z$ axis
and labeled from $1$ to $L$ (see Fig.~\ref{fig:schema2_FSF}).
Two consecutive layers $n$
and $n+1$ are coupled by a tunnel amplitude $t_n$.
We use Green's functions that are parametrized by the wave vector
in the $(x,y)$ direction and by the spatial coordinate in the 
lateral direction.

We note $\Hh_{i,j}^{(L)}$ the Green's functions of
the system of $L$ layers in a given electrode
and $\Hg_{i,i}$ the Green's function
of the isolated layer number $i$. The Green's function
$\Hh_{L,L}^{(L)}$ can be calculated recursively through
a matrix continued fraction:
\be
\label{eq:matrix}
\Hh_{L,L}^{(L)}=\left[ \HI - \Hg_{L,L}
\Ht_{L-1} \Hh_{L-1,L-1}^{(L-1)} \Ht_{L-1} \right]^{-1}
\Hg_{L,L}
.
\ee
The local Green's functions of the system of $L$ stacked
layers are obtained through 
\be
\Hh_{i,i}^{(L)}=\Hh_{i,i}^{(L-1)}
+\Hh_{i,L-1}^{(L-1)} \Ht_{L-1} \Hh_{L,L}^{(L)}
\Ht_{L-1} \Hh_{L-1,i}^{(L-1)}
,
\ee
where $\Hh_{i,L}^{(L)}$ and $\Hh_{L,i}^{(L)}$ are
calculated recursively through the relations
$\Hh_{L,i}^{(L)} = \Hh_{L,L}^{(L)} \Ht_{L-1} \Hh_{L-1,i}^{(L-1)}$
and 
$\Hh_{i,L}^{(L)} = \Hh_{i,L-1}^{(L-1)} \Ht_{L-1} \Hh_{L,L}^{(L)}$.
The computation time required to obtain $\hat{h}_{L,L}^{(L)}$ 
is proportional to $L$ whereas it is proportional to $L^2$ if one
wants to calculate all the $\hat{h}_{i,i}^{(L)}$.

\subsection{Green's functions of the connected trilayer}

We note $\HG$ the
Green's functions of the connected F/S/F trilayer
(see Fig.~\ref{fig:schema2_FSF}).
Layers $a$ and $\alpha$ are
connected by a tunnel amplitude $t_{a,\alpha}=t_{\alpha,a}$
and layers $b$ and $\beta$ are connected by a tunnel
amplitude $t_{b,\beta}=t_{\beta,b}$.
We use the notation $t=t_{a,\alpha}=t_{b,\beta}$ and denote
by $t_0$ the
tunnel amplitude within each layer (see Eq.~\ref{eq:tunnel-t0}
and Eq.~\ref{eq:tunnel-t}).

The definition of the tunnel Hamiltonian that we use for the
F/S/F trilayer is slightly different from the conventional
definition given by (\ref{eq:tunnel-t}):
\begin{eqnarray}
{\cal W}_{\rm FSF}&=& \frac{t}{\sqrt{2}} \sum_{{\bf x},\sigma}
\left( c_{x,y,\alpha,\sigma}^+ c_{x,y,a,\sigma} +
c_{x,y,a,\sigma}^+ c_{x,y,\alpha,\sigma} \right)\\\nonb
&+& \frac{t}{\sqrt{2}} \sum_{{\bf x},\sigma}
\left( c_{x,y,\beta,\sigma}^+ c_{x,y,b,\sigma} +
c_{x,y,b,\sigma}^+ c_{x,y,\beta,\sigma} \right)
.
\end{eqnarray}
With the factors $\sqrt{2}$ 
the F/S/F trilayer with $L_S/a_0=1$ in the parallel alignment
is equivalent to the F/S bilayer
with $L_S/a_0=1$ and a tunnel amplitude equal to $t$. 
The correspondence between the bi and trilayer is useful
for checking the numerical simulations.
For $L_S/a_0 \ge 2$ we carry out the
simulations of the F/S/F trilayer but we note that 
in the parallel alignment
the qualitative predictions are valid also for F/S bilayers
as long as the thickness of the superconductor is smaller
than the coherence length.

We note
$\HK_{\alpha,\alpha}=\Hh_{\alpha,\alpha} \Ht_{\alpha,a}
\Hh_{a,a} \Ht_{a,\alpha}$,
$\HK_{\beta,\beta}=\Hh_{\beta,\beta} \Ht_{\beta,b}
\Hh_{b,b} \Ht_{b,\beta}$,
$\HK_{\alpha,\beta}=\Hh_{\alpha,\beta} \Ht_{\beta,b}
\Hh_{b,b} \Ht_{b,\beta}$,
$\HK_{\beta,\alpha}=\Hh_{\beta,\alpha} \Ht_{\alpha,a}
\Hh_{a,a} \Ht_{a,\alpha}$.
The Green's function $\HG_{\alpha,\alpha}$ is given by
\begin{eqnarray}
\HG_{\alpha,\alpha} &=& \left[ \HI - \HK_{\alpha,\alpha}
-\HK_{\alpha,\beta} \left[\HI-\HK_{\beta,\beta} \right]^{-1}
\HK_{\beta,\alpha} \right]^{-1}\\
&& \times
\left[ \Hh_{\alpha,\alpha} + \HK_{\alpha,\beta}
\left[\HI-\HK_{\beta,\beta}\right]^{-1} \Hh_{\beta,\alpha} \right]
,
\nonb
\end{eqnarray}
and $\HG_{\beta,\alpha}$ is given by
\be
\HG_{\beta,\alpha} = \left[\HI-\HK_{\beta,\beta} \right]^{-1}
\left[ \Hh_{\beta,\alpha} + \HK_{\beta,\alpha} \HG_{\alpha,\alpha}
\right]
.
\ee
The Green's function $\HG_{\alpha,\beta}$ is deduced from
$\HG_{\beta,\alpha}$ through the relation
$\HG_{\alpha,\beta}^{\tau_1,\tau_2} = \HG_{\beta,\alpha}^{\tau_2,\tau_1}$,
where $\tau_1$ and $\tau_2$ are the Nambu indexes.
$\HG_{\beta,\beta}$ is given by
\be
\HG_{\beta,\beta} = \left[\HI-\HK_{\beta,\beta}\right]^{-1}
\left[\Hh_{\beta,\beta}+\HK_{\beta,\alpha} \HG_{\alpha,\beta} \right]
.
\ee
We deduce the values of $\HG_{a,a}$, $\HG_{b,b}$, $\HG_{a,b}$ and $\HG_{b,a}$ 
as well as $G_{i,i}$ in the superconductor:
\begin{eqnarray}
\nonb
\HG_{i,i} &=& \Hh_{i,i}
+ \Hh_{i,\alpha} \Ht_{\alpha,a} \HG_{a,a} \Ht_{a,\alpha} \Hh_{\alpha,i} 
+ \Hh_{i,\alpha} \Ht_{\alpha,a} \HG_{a,b} \Ht_{b,\beta} \Hh_{\beta,i} \\
&+& \Hh_{i,\beta} \Ht_{\beta,b} \HG_{b,a} \Ht_{a,\alpha} \Hh_{\alpha,i} 
+ \Hh_{i,\beta} \Ht_{\beta,b} \HG_{b,b} \Ht_{b,\beta} \Hh_{\beta,i} 
.
\end{eqnarray}
To obtain the pair amplitude in the
ferromagnets and superconductor we first
calculate recursively the $\hat{h}$'s and next evaluate $\HG_{1,2}$.
The evaluation of the self-consistent superconducting gap is done
either by dichotomy if $L_S/a_0=1,2$ or by iterations
of the self-consistency equation (\ref{eq:self})
if $L_S/a_0 \ge 3$.

\section{F/S/F trilayers with $L_S,L_F \ll \xi_S^{(\perp)},\xi_F^{(\perp)}$}
\label{sec:FSFtri}
\label{sec:small}

In this section we consider the regime $L_S,L_F \ll \xi_S^{(\perp)},\xi_F^{(\perp)}$,
establish a connection between the LDOS and the self-consistent superconducting gap,
and show that the regime $h_{\rm ex}/\Delta_0\sim 1$ is characterized by 
Andreev bound states competing with the formation of a minigap.

\subsection{Self-consistent superconducting gap for weak ferromagnets
in the parallel alignment}
\label{sec:self}

\subsubsection{Role of the thicknesses of the electrodes}
\label{sec:role-thickness}

\begin{figure}
\includegraphics [width=.9 \linewidth]{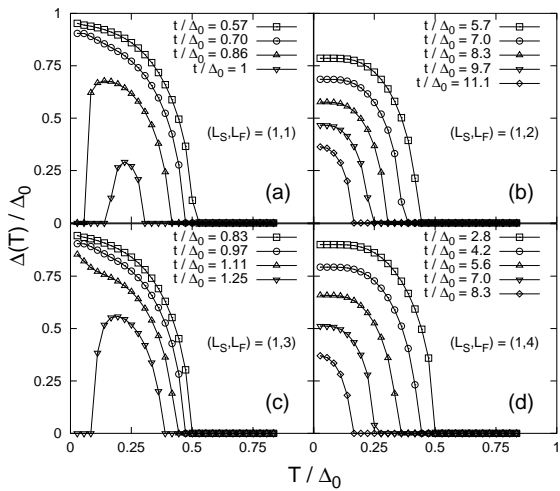}
\caption{Temperature dependence of the self-consistent
superconducting gap for 
the F/S/F trilayer in the parallel alignment with
$(L_S/a_0,L_F/a_0)=(1,1)$ (a), $(1,2)$ (b), $(1,3)$ (c), $(1,4)$ (d),
with weak ferromagnets ($h_{\rm ex} /\Delta_0 = 0.83$).
$L_F$ and $L_S$ are small compared to
$\xi_{S,0}^{(\perp)}=2 t_0 a_0/ \pi \Delta_0 = 17.7 a_0$,
$\xi_F^{(\perp)}=2 t_0 a_0 / \pi h_{\rm ex} =21.2 a_0$
and $\lambda_F=6.28 a_0$.
We use $\Delta_0/\epsilon_F=0.014$ and $t_0 / \epsilon_F=0.4$.
\label{fig:Delta1}
}
\end{figure}

We have shown on Fig.~\ref{fig:Delta1} the temperature dependence of the
superconducting gap for values of $(L_S,L_F)$ such that
$L_S,L_F \ll \xi_S^{(\perp)},\xi_F^{(\perp)}$. We used $L_S/a_0=1$ on
Fig.~\ref{fig:Delta1} but similar results were obtained with
$L_S/a_0=2$.
As expected
from Appendix~\ref{app:reso} the breakdown of
superconductivity in the resonant F/S/F trilayer
in the parallel alignment resembles
the case $(L_S/a_0,L_F/a_0)=(1,1)$\cite{Melin-Feinberg-FSF}:
as temperature is reduced the superconducting gap increases,
reaches a maximum and decreases to zero.
We obtain a reentrant behavior
in a narrow range of interface transparencies (not shown on
Fig.~\ref{fig:Delta1}). By contrast we obtain a monotonic
behavior for off-resonant values of $(L_S/a_0,L_F/a_0)$, as expected
from Appendix~\ref{app:reso}.

The values of the tunnel amplitude needed to destroy superconductivity
for off-resonant trilayers (see Figs.~\ref{fig:Delta1}-(b) and (d))
is almost ten times larger than in the resonant case
(see Figs.~\ref{fig:Delta1}-(a) and (c)). This is because the effective
coupling between the superconductor and ferromagnets is much weaker
in the off-resonant case due to the differences in the density of states.

\subsubsection{Role of the exchange field}

We repeated the simulations
for the F/S/F trilayer with
$(L_S/a_0,L_F/a_0)=(2,2)$ but with different values of $h_{\rm ex}/\Delta_0$.
For the smallest value of $h_{\rm ex}/\Delta_0$ ($h_{\rm ex}/\Delta_0=0.28$)
we obtained a monotonic decrease of the superconducting gap
as a function of temperature for all values of the tunnel amplitude.
For $h_{\rm ex}/\Delta_0=0.56, 0.83, 1.11, 1.39$ we obtained a
non monotonic temperature dependence of the superconducting gap
similar to Fig.~\ref{fig:Delta1}-(a) and Fig.~\ref{fig:Delta1}-(b).
The non monotonic
variation of the superconducting gap 
occurs typically for $h_{\rm ex}$ being a fraction of $\Delta_0$
up to values slightly above $\Delta_0$.

\subsection{Relation between the local density of states and
the self-consistent superconducting gap}

\subsubsection{Local density of states in the parallel alignment}
\label{sec:peak}
\label{sec:parallel}

\begin{figure}
\includegraphics [width=1. \linewidth]{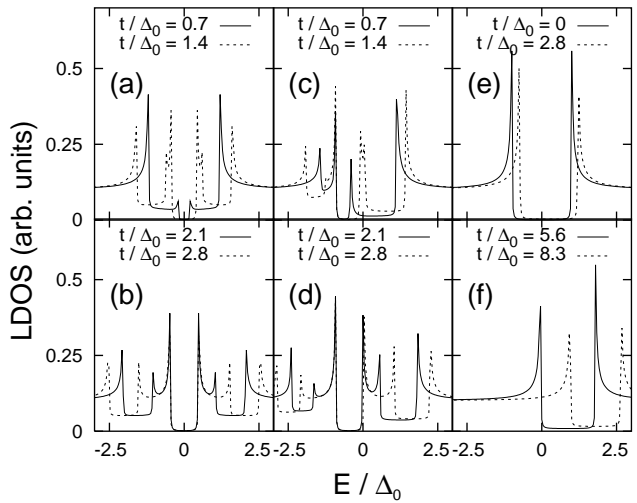}
\caption{Energy dependence of the spin-up LDOS
(in arbitrary units)
in the superconducting layer of the F/S/F
trilayer in the parallel alignment with
$(L_S/a_0,L_F/a_0)=(1,3)$. The
superconducting gap, equal to $\Delta_0$, is not self-consistent.
(a) and (b) correspond to $h_{\rm ex}/\Delta_0=0$ , $\xi_F^{(\perp)}=\infty$
(N/S interface).
(c) and (d) correspond to $h_{\rm ex}/\Delta_0=0.83$, $\xi_F^{(\perp)}=21.2 a_0$
(F/S interface with
weak ferromagnets).
(e) and (f) correspond to $h_{\rm ex}/\Delta_0=13.9$, $\xi_F^{(\perp)}=1.3 a_0$
(F/S interface with strong ferromagnets).
We use $\Delta_0/\epsilon_F=0.014$. $t_0 / \epsilon_F=0.4$,
$\eta/\Delta_0=8.3 \times 10^{-3}$, $\lambda_F=6.28 a_0$.
\label{fig:LDOS1}
}
\end{figure}

The spin-up LDOS $\rho_\uparrow(\omega)$
in the superconductor is shown on Fig.~\ref{fig:LDOS1} for
the F/S/F trilayer in the parallel alignment, for
$(L_S/a_0,L_F/a_0)=(1,3)$. We obtained similar results for
$(L_S/a_0,L_F/a_0)=(2,2)$.
The spin-down LDOS $\rho_\downarrow(\omega)$ is obtained through
the relation $\rho_\downarrow(\omega)=\rho_\uparrow(-\omega)$.
The zero temperature superconducting gap $\Delta$
is fixed to the BCS value $\Delta_0$ of an isolated superconducting
layer.

The case of a N/S interface is shown on Fig.~\ref{fig:LDOS1}-(a) and
Fig.~\ref{fig:LDOS1}-(b).
The LDOS is symmetric with respect to a
change of sign in energy, as expected in the absence of an exchange field.
There exists a minigap at the Fermi energy so that superconductivity is
robust against increasing the tunnel amplitude $t$.
Increasing
$t$ gives rise to pairs of
Andreev bound states at opposite energies
(see Fig.~\ref{fig:LDOS1}-(a)). Each peak corresponds
to a miniband, which is visible on Fig.~\ref{fig:LDOS1}-(b)
obtained with larger values of $t$.
The formation of the minibands is related to the
infinite planar geometry: the layers are infinite in the $x$
and $y$ directions (see Fig.~\ref{fig:schema1_FSF})
and there is thus a degeneracy
associated to the position of the Andreev bound state in the $(x,y)$ plane.
The bound state can delocalize in the $(x,y)$ plane
and thus acquire a dispersion which would not occur if
the dimensions $H$ and $D$ (see Fig.~\ref{fig:schema1_FSF}) were small
compared to the BCS coherence length, a situation considered
in Ref.~\onlinecite{Vecino}.

For weak ferromagnets with $t$, $h_{\rm ex}$ and $\Delta$ having
the same order of magnitude (panel (c) on Fig.~\ref{fig:LDOS1})
we obtain one Andreev bound state miniband inside the gap
and one resonant scattering state above the gap.
The Andreev bound state miniband
moves to the Fermi energy as the tunnel amplitude $t$ increases
(see Fig.~\ref{fig:LDOS1}-(c)) and appears at a positive
energy for larger values of $t$ (see Fig.~\ref{fig:LDOS1}-(d)).

For larger
values of the exchange field the Andreev bound state miniband
disappears from the LDOS. For $h_{\rm ex}/\Delta_0 \gg 1$ 
the induced exchange field is opposite to the magnetization
in the ferromagnets\cite{Jirari,Volkov}.

There are thus two qualitatively different depairing mechanisms for
strong and weak ferromagnets.
For strong ferromagnets with $h_{\rm ex}/\Delta_0 \gg 1$ (see Figs.~\ref{fig:LDOS1}-(e)
and~\ref{fig:LDOS1}-(f))
the breakdown
of superconductivity is due mainly to Zeeman splitting. 
The case of very weak ferromagnets ($h_{\rm ex} /\Delta_0\ll 1$) 
resembles that case $h_{\rm ex}/\Delta_0=0$: non perturbative Andreev bound states
are generated in the superconducting gap approximately at opposite energies.
In the case $h_{\rm ex}/\Delta_0 \sim 1$ the Andreev bound state miniband can be
at zero energy, therefore destructing the
minigap.

\subsubsection{Finite temperature local density of states}

\begin{figure}
\includegraphics [width=.8 \linewidth]{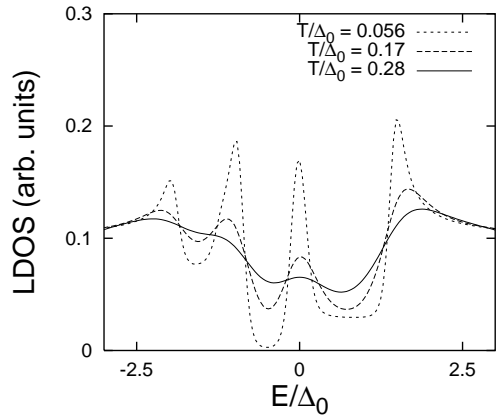}
\caption{Energy dependence of the
finite temperature spin-up LDOS (in arbitrary units)
in the superconducting layer of the F/S/F
trilayer in the parallel alignment with $(L_S/a_0,L_F/a_0)=(1,3)$.
The zero temperature LDOS corresponds to Fig.~\ref{fig:LDOS1}-(c)
with $t/\Delta_0=1.4$.
\label{fig:T}
}
\end{figure}

The spin-up LDOS at a finite temperature $T$ is related the conductance
of a scanning tunneling microscope (STM)
in which the tip is made of a half-metal ferromagnet.
The finite temperature current at an arbitrary voltage is obtained
through Keldysh formalism.
The finite temperature LDOS is found to be
\be
\label{eq:LDOS-Temp}
\rho_T(\omega)=\int d \omega' \frac{\rho(\omega')}{4 T
\cosh^2{\left( \frac{\omega'-\omega}{2 T} \right)}}
,
\ee
where $\rho(\omega)$ is equal to the zero temperature LDOS.

We calculated the finite temperature LDOS of the $(L_S/a_0,L_F/a_0)=(1,3)$
trilayer in the parallel alignment
(see Fig.~\ref{fig:T}).
Increasing temperature tends
to reduce the intensity of the Andreev bound state
at the Fermi energy on Fig.~\ref{fig:LDOS1}-(c)
so that the LDOS at the Fermi energy decreases if $T$ increases.
For relatively large values of $T$
the peak structure has almost disappeared from the
LDOS but there remains a minimum associated
to the superconducting gap that is
significantly filled.
As $T$ increases there is thus a 
first cross-over
where the peak at the Fermi energy disappears and
a superconducting minigap is restored, and a second
cross-over where the superconducting gap
disappears. The first cross over occurs at a temperature
equal to the bandwidth of the Andreev bound state miniband
and the second cross-over occurs at a temperature comparable to
the zero temperature superconducting gap.
The finite temperature LDOS is thus in a qualitative agreement with
the non monotonic self-consistent superconducting gap.

Reentrance obtained in a narrow
range of interface transparencies\cite{Melin-Feinberg-FSF} can
also be explained by this qualitative picture: if the Andreev
bound state miniband
is narrow and located at a slightly positive energy
like on Fig.~\ref{fig:LDOS1}-(c) then the LDOS
at the Fermi energy
is vanishingly small at $T=0$
since the Fermi energy is not
in the Andreev bound state miniband. By increasing $T$ the width
of the Andreev bound state miniband increases so that the
density of states at the Fermi energy increases. By further increasing $T$ 
the intensity of the Andreev bound state
miniband is reduced, and the density of states
at the Fermi energy decreases.
This behavior is compatible with a reentrant
behavior of the self-consistent superconducting gap.
The correlation between the low temperature superconducting gap
and the LDOS at the Fermi energy
is further established in section~\ref{sec:corre}.

\subsubsection{Local density of states in the parallel alignment}
\label{sec:qual-AP}

\begin{figure}
\includegraphics [width=1. \linewidth]{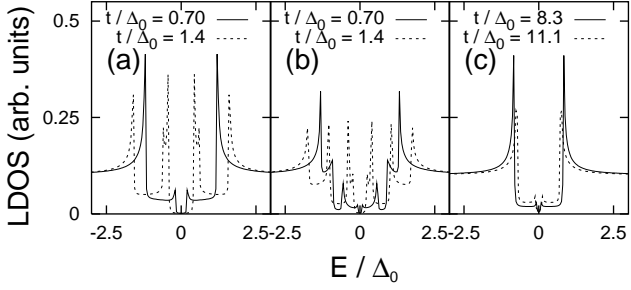}
\caption{Energy dependence of the spin-up LDOS
(in arbitrary units)
in the superconducting layer of the F/S/F
trilayer in the antiparallel alignment with
$(L_S/a_0,L_F/a_0)=(1,3)$.
The superconducting gap, equal to $\Delta_0$, is not self-consistent.
(a) and (b) correspond to $h_{\rm ex} / \Delta_0=0.83$,
$\xi_F^{(\perp)}=21.1 a_0$
(weak ferromagnets).
(c) corresponds to $h_{\rm ex}/\Delta_0=2.78$,
$\xi_F^{(\perp)}=6.4 a_0$ (strong ferromagnets).
We used
$\Delta_0/\epsilon_F=0.014$, $t_0 / \epsilon_F=0.4$,
$\eta/\Delta_0=8.3 \times 10^{-3}$
and $\lambda_F=6.28 a_0$.
\label{fig:LDOS2}
\label{fig:anti}
}
\end{figure}

The spin-up LDOS in the antiparallel alignment
is shown on Fig.~\ref{fig:LDOS2}.
The
LDOS is symmetric with respect to a change of sign of energy,
but not equivalent to the LDOS of a N/S interface
(see Fig.~\ref{fig:LDOS1}-(a) and Fig.~\ref{fig:LDOS1}-(b)).
There exists a well-defined minimum at the Fermi energy
corresponding to the superconducting minigap.
The energy dependence
of the LDOS in the antiparallel alignment shows that 
superconductivity is stronger than in the parallel alignment,
both for weak and strong ferromagnets. This is expected
because of the
exchange field induced in the superconductor in
the antiparallel alignment is reduced compared to the
parallel alignment~\cite{deGennes-FSF}.
The self-consistent superconducting gap has a monotonic temperature
dependence in this case which is because a well-defined minigap is
obtained in the LDOS.

\subsection{Role of disorder for weak ferromagnets and for the
F/S/F trilayer with atomic thickness}
\label{sec:dis}

Disorder plays an important role in the proximity effect. Diffusive N/S interfaces
are described by quasiclassical theory\cite{Belzig-quasi}. 
A small disorder can be incorporated in our description
based on microscopic Green's functions like in Ref.~\onlinecite{Abrikosov}
(see Appendix~\ref{app:disorder}).
The strength of disorder in the superconductor is characterized
by $\delta_S^{(0)}= \sqrt{n_\alpha u_\alpha^2}/\Delta_0$,
where $n_\alpha$ is the concentration
of impurities and $u_\alpha$ is the scattering potential,
and we use a similar parameter
$\delta_F^{(0)}$
to characterize disorder in the ferromagnets. We obtain a significant effect
of disorder for
relatively large values of $\delta_S^{(0)}$ and $\delta_F^{(0)}$.

\subsubsection{Disorder in the superconducting and ferromagnetic layers}
\label{sec:dis1}

\begin{figure}
\includegraphics [width=1. \linewidth]{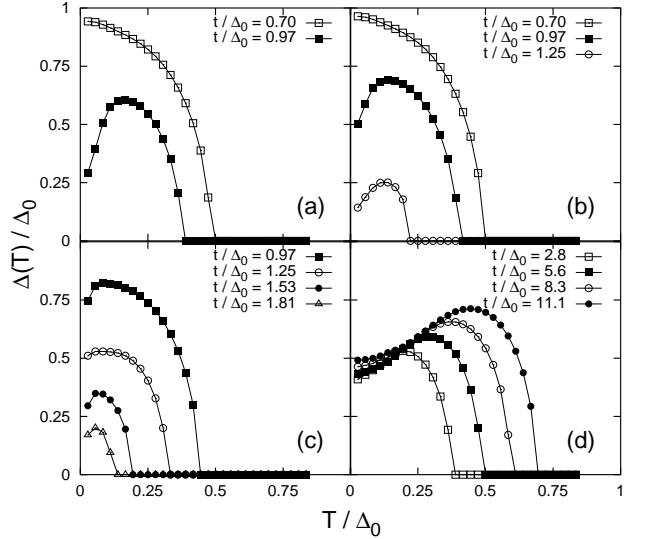}
\caption{Temperature dependence of the self-consistent superconducting gap
of the F/S/F trilayer in the parallel alignment with $(L_S/a_0,L_F/a_0)=(1,1)$
in the presence of disorder in the superconducting layer.
Without vertex corrections we use
$h_{\rm ex} /\Delta_0= 0.83$,
$\delta_S^{(0)}=6.2$ (a), $\delta_S^{(0)}=7.6$ (b),
$\delta_S^{(0)}=9.8$ (c).
(d) corresponds to $\delta_S^{(0)}=9.8$ with vertex corrections.
In all cases we use $\delta_F^{(0)}=0$.
\label{fig:dis1}
}
\end{figure}

The temperature dependence of the self-consistent superconducting gap in
the presence of disorder in the superconductor is shown
on Fig.~\ref{fig:dis1} for $(L_S/a_0,L_F/a_0)=(1,1)$
in the parallel alignment. In the absence of disorder Fig.~\ref{fig:dis1}
corresponds to Fig.~\ref{fig:Delta1}-(a).
The effect of disorder is to reduce the effect of 
the tunnel amplitude\cite{Melin-Feinberg-FSF}
so that if disorder increases a larger value of $t$
is needed to destroy superconductivity.
Comparing Fig.~\ref{fig:Delta1} and Fig.~\ref{fig:dis1}
we see that the variations of $\Delta(T)$ are affected by
a weak disorder especially if the
F/S/F trilayer in the parallel alignment is close to the breakdown of
superconductivity.
In this case the dimensionless parameter controlling
the strength of disorder is $\delta_S=\delta_S^{(0)} \Delta_0/\Delta(T)$
which can be much larger than $\delta_S^{(0)}$.

\subsubsection{Vertex corrections}
\label{sec:vertex1}
There exist two perturbative series: one in the
hopping amplitude $t$
and the other in the disorder scattering potential $u$.
Vertex corrections arise from diagrams that mix the two series
(see Appendix~\ref{app:dis-vertex}).
The temperature dependence of the self-consistent
superconducting gap with the vertex corrections
is shown on Fig.~\ref{fig:dis1}-(d).
The critical temperature is larger
if vertex corrections are included. We
obtain a non monotonic temperature dependence of the
superconducting gap even in the presence of vertex corrections.
The role of vertex corrections increases if $t$ increases
since the vertex correction term is proportional to
$n_\alpha t^2 u_\alpha^2$.

\section{Finite thickness in the superconductor
($L_S \alt \xi_S^{(\perp)}$ and $L_F \ll \xi_F^{(\perp)}$)}
\label{sec:finite}
\label{sec:VC}

\begin{figure}
\includegraphics [width=1. \linewidth]{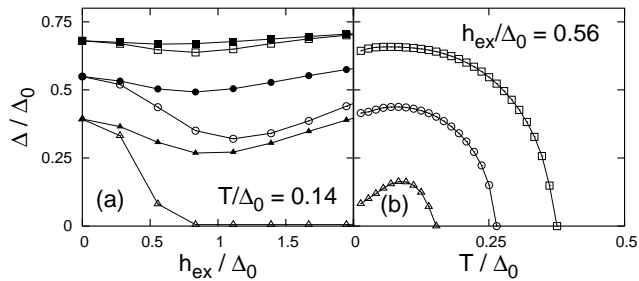}
\caption{
(a) Variation of the self-consistent superconducting gap
calculated at $T/\Delta_0=0.14$ in the parallel (P)
alignment (open symbols) and in the antiparallel (AP) alignment
(filled symbols) as a function of $h_{\rm ex}/\Delta_0$.
(b) Variation of the self-consistent
superconducting gap calculated with
$h_{\rm ex}/\Delta_0=0.56$ in the parallel alignment, as
a function of $T/\Delta_0$.
We used in both cases $(L_F/a_0,L_S/a_0)=(11,11)$. The Fermi wave length
is $\lambda_F=6.28 a_0$.
We use
$t/\Delta_0=5.6$ ($\square$ in the
P alignment and $\blacksquare$ in the AP
alignment),
$t/\Delta_0=8.3$ ($\circ$ in the P alignment and
$\bullet$ in the AP alignment), and $t/\Delta_0=11.1$
($\triangle$ in the P alignment and $\blacktriangle$
in the AP alignment).
\label{fig:P-AP}
}
\end{figure}

In this section we include a finite thickness in the superconductor for
$L_F \ll \xi_F^{(\perp)}$. For weak ferromagnets in the regime
$h_{\rm ex}/\Delta_0\sim1$ we obtain non monotonic temperature dependences
of the self-consistent superconducting gap in the parallel
alignment, therefore confirming section~\ref{sec:small}.

We have shown on Fig.~\ref{fig:P-AP}-(a) the variation of the self-consistent
superconducting gap in the middle of the superconductor as a function
of the reduced exchange field $h_{\rm ex}/\Delta_0$.
As expected from the variation of the critical
temperature\cite{Baladie,Houzet,Fominov} we obtain a minimum if the exchange
field is comparable to the superconducting gap. Moreover we obtain
a minimum also in the antiparallel alignment that is related to the
minigap formed in between Andreev bound states at opposite energies
(see section~\ref{sec:qual-AP}).
The full temperature
dependence of the self-consistent superconducting gap in the parallel
alignment is shown on Fig.~\ref{fig:P-AP}-(b) for $L_S/a_0=11$
(comparable to $\xi_S^{(\perp)}/a_0$) and $L_F/a_0=11$ (much smaller than
$\xi_F^{(\perp)}/a_0$). We obtain a maximum in the variation of
$\Delta(T)/\Delta_0$ in the parallel alignment.
We carried out the
same simulation in the antiparallel alignment and found a monotonic
temperature dependence of the self-consistent superconducting gap
(not shown on Fig.~\ref{fig:P-AP}).

\section{Finite thickness in the ferromagnets
($L_S \ll \xi_S^{(\perp)}$ and $L_F \agt \xi_F^{(\perp)}$)}
\label{section:pi}
\label{sec:corre}

We discuss now the regime $L_S\ll\xi_S^{(\perp)}$ and $L_F\agt \xi_F^{(\perp)}$.
We find Andreev bound states at the Fermi energy in the parallel alignment for
strong ferromagnets, correlated with non monotonic temperature dependences of the
self-consistent superconducting gap.

\begin{figure}
\includegraphics [width=.9 \linewidth]{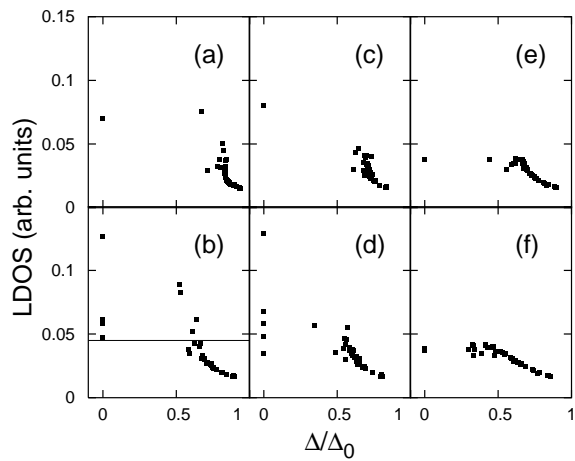}
\caption{Correlation between $\Delta/\Delta_0$ and the LDOS
at the Fermi energy (in arbitrary units)
for $L_S/a0=1$ and $L_F/a_0$ between $1$ and $40$. The temperature
is $T/\Delta_0=0.14$.
In the calculation of the LDOS the superconducting gap is not self-consistent.
We use strong ferromagnets with $h_{\rm ex}/\Delta_0=13.9$ and
$\xi_F^{(\perp)}=1.3 a_0$.
We use $L_S/a_0=1$ and $t/\Delta_0=2.1$ (a);
$L_S/a_0=1$ and $t/\Delta_0=2.8$ (b);
$L_S/a_0=2$ and $t/\Delta_0=3.5$ (c);
$L_S/a_0=2$ and $t/\Delta_0=4.2$ (d);
$L_S/a_0=1$ and $t/\Delta_0=2.8$ (e);
$L_S/a_0=1$ and $t/\Delta_0=3.3$ (f).
(a), (b), (c), (d) correspond to the parallel alignment.
(e) and (f) correspond to the antiparallel alignment.
The solid line in (b) corresponds to $\rho(\omega=0)=0.045$
(see text).
\label{fig:corre}
}
\end{figure}

The regime $L_F \agt \xi_F$ is
characterized by oscillations of the self-consistent superconducting gap,
critical temperature and pair amplitude as a function of $L_F$.
We obtain bound states within the superconducting gap
and resonant scattering states outside the superconducting gap.
We calculated
systematically the LDOS and the self-consistent
superconducting gap
for $L_S/a_0=1,2$ and
$L_F/a_0=1,...,100$, in the parallel and antiparallel alignments
(see Fig.~\ref{fig:corre}). The systematic calculation of the finite temperature
LDOS given by Eq.~(\ref{eq:LDOS-Temp}) is too demanding from a computational point of
view. Instead we calculate the LDOS at zero temperature with $\eta=T$.
The consistency between the two calculations was verified in a few cases.
Depending on the
interface transparencies $\Delta(L_F/a_0)/\Delta_0$ 
tends either to a finite value or to zero
in the limit $L_F/a_0 \rightarrow + \infty$.
We concentrate on the first case only.
For $1 \le L_F/a_0 \le 40$ we
see on Fig.~\ref{fig:corre}
that on average the superconducting gap is smaller when
the LDOS at the Fermi energy is larger. For $L_S/a_0=1$
and $t/\Delta_0=2.8$ (see Fig.~\ref{fig:corre}-(b)) we obtain 
$\Delta(T)/\Delta_0=0$ 
for four values of $L_F/a_0$ ($L_F/a_0=4,6,11,13$).
For three other values of $L_F/a_0$ ($L_F/a_0=16,23,25$)
we obtain a non monotonic variation of $\Delta(T)/\Delta_0$.
For two other values of $L_F/a_0$ ($L_F/a_0=32$ and $L_F/a_0=35$)
$\Delta(T)/\Delta_0$ is monotonic
but far from the BCS variation. These nine values of $L_F/a_0$ 
with an anomalous temperature dependence of the self-consistent
superconducting gap have
large values of the LDOS at the Fermi energy ($\rho(\omega=0)>0.045$
on Fig.~\ref{fig:corre}-(b), in arbitrary units).
Like in section~\ref{sec:parallel}
Andreev bound states near the Fermi
energy correlate with unconventional temperature dependences of
the superconducting gap.

For $40 \le L_F/a_0 \le 100$ corresponding to $L_F \gg \xi_F^{(\perp)}$,
we obtain a cloud of points with a small dispersion and with 
no correlation between the superconducting gap and the LDOS at 
the Fermi
energy. This corresponds to the cross-over to the
F/S/F trilayer with bulk ferromagnets.

We obtain bound states 
within the superconducting gap and unconventional temperature
dependences of the superconducting gap
with strong ferromagnets in the parallel alignment. By contrast
with strong ferromagnets we obtain a conventional temperature
dependence of the superconducting gap for the F/S/F trilayer
with smaller values of $(L_S/a_0,L_F/a_0)$\cite{Melin-Feinberg-FSF}.

We carried out the same simulation in the antiparallel alignment
and found that $\Delta(T)/\Delta_0$ is close to the BCS 
temperature dependence of the superconducting gap
for all values of $L_S/a_0$ between $1$ and $100$.
A larger density of states at the Fermi energy correlates with
smaller values of the self-consistent superconducting gap
like in the parallel alignment (see Fig.~\ref{fig:corre}-(e) and (f)).
Because of the minigap that can be formed since
the bound states are at opposite energies (see Fig.~\ref{fig:anti})
we do not obtain the points with large values of the density of
states at the Fermi energy like in the parallel alignment.
The absence of large values of the LDOS at the Fermi
energy is related
to the fact that the Andreev levels in the antiparallel alignment
do not cross the Fermi energy as the interface transparency is increased.

\section{Comparison with other models}
\label{sec:compa}
The discussion in the preceding sections was restricted
to the infinite planar geometry.
Now we discuss Andreev bound states in
other geometries without imposing self-consistency on
the superconducting gap.
In section~\ref{sec:1D} we
improve the discussion of a model proposed by
recently Vecino {\sl et al.}\cite{Vecino}
in which a ferromagnetic chain
is connected to a superconductor\cite{note}.
In section~\ref{sec:Cserti} we compare
the LDOS
to a model discussed recently by Cserti {\sl et al.}\cite{Cserti}.

\subsection{A one dimensional model}
\label{sec:1D}

\begin{figure}
\includegraphics [width=1. \linewidth]{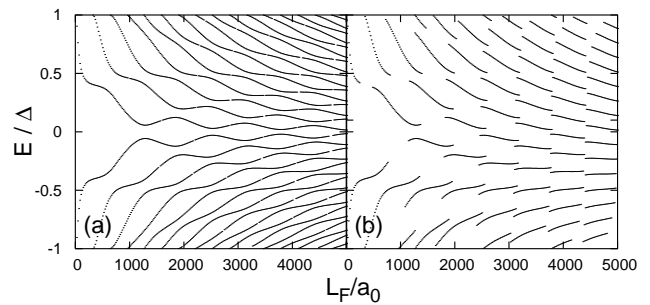}   
\caption{Reduced energy $E/\Delta$
of the Andreev bound states as a function of
the number of sites $L_F/a_0$ in the ferromagnetic chain for
weak ferromagnets ($h_{\rm ex}/\Delta=0.5$).
We used the parameters $\Delta/t=\Delta_0/t_F=0.01$,
$t/t_S=0.01$ , $L_S=10^5 a_0$.
The entire spectrum is kept in (a). Only the levels with 
a residue larger than $2$ are kept in (b). The superconducting gap
is not self-consistent.
\label{fig:ABS}
}
\end{figure}

We suppose that the
superconductor and ferromagnets are described by 1D
chains with open boundary conditions with
$L_S/a_0$ sites in the superconductor and
$L_F/a_0$ sites in the ferromagnet.
We denote by $t_S$ ($t_F$)
the hopping amplitude in the superconductor (ferromagnet)
and we use $t_S=t_F$.
The energy level spacing in the superconductor is much smaller
than the superconducting gap $\Delta$.

We have shown on Fig.~\ref{fig:ABS} the evolution of the
energy of Andreev bound states as a function of the length
of the ferromagnetic chain obtained with the algorithm presented
in Appendix~\ref{app:algo}.
The spectrum is symmetric under a change of sign in energy
if we keep all energy levels into account.
Andreev bound states arising from the gap edges move to
the Fermi energy as $L_F/a_0$ increases\cite{Vecino}. There is
level repulsion between Andreev bound state and oscillations
of the energy levels with a period of order $10$ times
$\xi_F=2 t_F a_0/ \pi h_{\rm ex}$. These oscillations 
do not exist in Ref.~\onlinecite{Vecino}.
Among all Andreev bound states obtained at a fixed $L_F/a_0$ some have a
spectral weight much larger than the other.
We keep only the levels having a spectral weight
larger than a given cut-off. The evolution of
the remaining
Andreev bound states as a function of $L_F/a_0$ is then in
agreement with Ref.~\onlinecite{Vecino}. 

In connection with the discussion in section~\ref{sec:peak}
we see that Andreev bound states near the Fermi energy occur only
if the ferromagnetic chain is long enough, larger than
approximately $10$ times $\xi_S=2 t_S a_0/\pi\Delta$.
This should be
contrasted
with the LDOS in the infinite planar limit discussed
in section~\ref{sec:peak}, and with the
model discussed in section~\ref{sec:Cserti}.

\subsection{Bogoliubov-de Gennes equations}
\label{sec:2D}
\label{sec:Cserti}

\begin{figure}
\includegraphics [width=1 \linewidth]{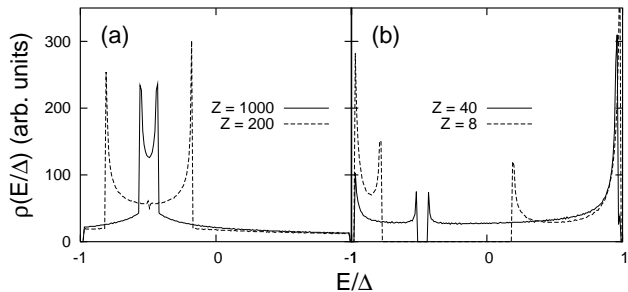}
\caption{Density of Andreev bound states (in arbitrary units)
$\rho(E/\Delta)$ as
a function of reduced energy $E/\Delta$ for $Z=1000$ and $Z=200$ (a);
$Z=40$ and $Z=8$ (b). We used the parameters $\Delta/\epsilon_F=0.02$,
$\Delta=0.01$, $h_{\rm ex}/\Delta=0.5$,
$d=\lambda_F/2$, $W\simeq 15900 \lambda_F$ where $\lambda_F=2 \pi / k_F$
is the Fermi wave length.
\label{fig:DOS}
}
\end{figure}

We consider now the model proposed by Cserti {\sl et al.}\cite{Cserti} 
in which a 2D ferromagnetic dot with a rectangular shape of dimensions
$(d,W)$ is connected to a superconductor
(see Fig.~\ref{fig:schema1_FSF}-(b)).
The superconductor has a width $W$ but is
infinite in the other direction ($d'=+\infty$ on
Fig.~\ref{fig:schema1_FSF}).
Using the solution of this model\cite{Cserti}
based on Bogoliubov-de Gennes equations
we generated the set $\{E_n\}$ of Andreev bound state energies.
Like for the de Gennes-Saint James
model\cite{deGennes-bound}
the density of bound states is not
equal to the LDOS in the superconductor\cite{Vecino}.
The evolution of the density of Andreev bound states as 
a function of the interface transparency is shown on
Fig.~\ref{fig:DOS}. The interface transparency is
parametrized by the dimensionless coefficient $Z$,
equal to the repulsive interface
potential in units of the Fermi energy\cite{BTK}.
For large values of $Z$ (corresponding to tunnel interfaces)
we obtain an Andreev bound state miniband around $E/\Delta_0=h_{\rm ex}/\Delta_0$.
As the interface transparency decreases the miniband broadens and
splits into two separate minibands (see Fig.~\ref{fig:DOS}).
This is in a qualitative agreement with section~\ref{sec:peak}
(see Fig.~\ref{fig:LDOS1}-(c) and Fig.~\ref{fig:LDOS1}-(d))
where we obtain also Andreev bound state minibands that evolve
inside the superconducting gap and can generate a large density
of states at the Fermi energy for some values of the 
interface transparencies.

\section{Conclusions}
\label{sec:conclu}

We have presented a detailed analysis of F/S/F trilayers with weak ferromagnets based
on microscopic Green's functions. 
The exchange field induced in the superconductor in the regime
$h_{\rm ex}/\Delta_0\gg1$ generates Zeeman splitting of the LDOS
the regime $h_{\rm ex}/\Delta_0\ll 1$ is characterized by Andreev bound states.
In the non perturbative regime $t/\Delta_0 \sim h_{\rm ex}/\Delta_0 \sim 1$ 
the Andreev bound state miniband crosses the Fermi energy
as the interface transparency
is increased therefore competing with the formation of a minigap,
which explains why the self-consistent superconducting gap is weakened
by the formation of an Andreev bound state miniband near the Fermi energy.
Increasing temperature  decreases the intensity of the Andreev bound state peak
in the LDOS which correlates with a reduction of the low temperature
self-consistent superconducting gap.

We found non monotonic temperature dependences of the superconducting gap
with weak ferromagnets
for $L_F,L_S \ll \xi_F^{(\perp)}, \xi_S^{(\perp)}$ as well as for
$L_S \sim \xi_S^{(\perp)}$ and $L_F \ll \xi_F^{(\perp)}$. In the regime
$L_S \ll \xi_S^{(\perp)}$ and $L_F \agt \xi_F^{(\perp)}$ we obtain non
monotonic temperature dependences for some values of $L_F/a_0$ for
strong ferromagnets.

For the F/S/F trilayer in the parallel alignment we find pairs of
Andreev bound states at opposite energies but there exists a minigap
so that the self-consistent superconducting gap decreases monotonically
with temperature. However at a fixed temperature the self-consistent
superconducting gap is non monotonic as a function of the exchange field
which is due to pairs of Andreev bound states at opposite energies in the
regime $h_{\rm ex}/\Delta_0 \sim 1 \ll t/\Delta_0$.

Concerning the induced exchange field in the F/S/F trilayer in the parallel
alignment we find that for weak
ferromagnets the exchange field is in the same direction as in the
ferromagnetic electrodes whereas it is in the opposite direction for
strong ferromagnets. The change of sign is visible both in the
LDOS and in the magnetization profile.
As the exchange field in the ferromagnets is increased we find a
first order transition to the normal state, as for paramagnetically limited
superconducting films in an applied magnetic field. 

\section*{ACKNOWLEDGMENTS}
The author wishes to thank D. Feinberg for fruitful discussions and
useful comments on the manuscript.

\appendix

\section{Resonances in lateral confinement}
\label{app:reso}
In this Appendix we account for the differences between the
resonant and off-resonant F/S/F trilayers.
The wave function in the lateral direction within a given
electrode is described by
the tight binding Hamiltonian
\be
{\cal H}=t_0 \sum_z \left[ | z+1 \rangle \langle z | 
+ |z \rangle \langle z+1 | \right]
,
\ee
where $t_0$ is the hopping between neighboring layers, and
$z/a_0$ is an integer between $0$ to $L/a_0-1$.
The eigenstates with open boundary conditions are given by
\be
|\psi_n\rangle = \sqrt{\frac{2 a_0}{L+a_0}}
\sum_{z=0}^{L/a_0-1} \sin{\left(
n \pi \frac{z+a_0}{L+a_0} \right)} | z \rangle
,
\ee
and the energy is given by
\be
\label{eq:epsilon-n}
\epsilon_n(L) =2 t_\perp \cos{\left(\frac{n \pi a_0}{L+a_0} \right)}
,
\ee
with $n=1, ... , L/a_0$. Let us consider a F/S/F trilayer with
$L_F/a_0$ layers in the ferromagnets and $L_S/a_0$
layers in the superconductor. 
Quasiparticles in the superconductor with a transverse
quantum number $n_S$ can tunnel in the
ferromagnets only if there exists an energy level
with quantum number $n_F$ close to resonance
in the ferromagnets, such that $\epsilon_{n_S}(L_S)
\simeq \epsilon_{n_F}(L_F)$.
If tunneling between the ferromagnets and superconductor is
not resonant then the F/S/F trilayer behaves as if the
ferromagnets were insulating.
The lowest off-resonant values of $(L_S/a_0,L_F/a_0)$ correspond to 
$(L_S/a_0,L_F/a_0)=(1,2)$, $(1,4)$, $(2,1)$, $(2,3)$, 
$(3,2)$, $(3,4)$, $(4,1)$, $(4,2)$, $(4,3)$.
The lowest resonant values of $(L_S/a_0,L_F/a_0)$ correspond to
$(L_S/a_0,L_F/a_0)=(1,1)$, $(1,3)$, $(2,2)$, $(3,1)$, $(3,3)$.

\section{Disorder in the F/S/F trilayer with atomic thickness}
\label{app:disorder}

In this Appendix we give the Dyson equations of a F/S/F trilayer
in the presence of disorder.

\subsection{Disorder in the superconducting and ferromagnetic layers}
We first neglect vertex corrections that
are discussed in section~\ref{sec:vertex1}.
We replace the Green's functions $\Hg_{\alpha,\alpha}$ of the superconducting layer
by the Green's function
$\Hg^{(d)}_{\alpha,\alpha}$ of the superconducting layer
in the presence of disorder. To second order in disorder the
Dyson equation takes the form
\begin{eqnarray}
\label{eq:Dyson-d}
&& \Hg^{(d)}_{\alpha,\alpha}({\bf k},\omega)
= \Hg_{\alpha,\alpha}({\bf k},\omega) 
+ n_\alpha a_0^2 \Hg_{\alpha,\alpha}({\bf k},\omega) \\\nonb
&\times& \int \frac{d {\bf k}'}{(2\pi)^2}
\overline{\Hu_\alpha({\bf k}-{\bf k}')
\Hg_{\alpha,\alpha}({\bf k}',\omega)
\Hu_\alpha({\bf k}'-{\bf k})}
\Hg^{(d)}_{\alpha,\alpha}({\bf k},\omega)
,
\end{eqnarray}
where $n_\alpha$ is the concentration of impurities
and the overline denotes an averaging over disorder.
Eq.~(\ref{eq:Dyson-d}) is solved according to
\begin{eqnarray}
&& \left[ \begin{array}{cc}
g_{\alpha,\alpha}^{(d),1,1} & f_{\alpha,\alpha}^{(d),1,2}\\
f_{\alpha,\alpha}^{(d),2,1} & g_{\alpha,\alpha}^{(d),2,2}
\end{array} \right] =
\frac{1}{\cal D} \left\{
\left[ \begin{array}{cc}
g_{\alpha,\alpha}^{1,1} & f_{\alpha,\alpha}^{1,2}\\
f_{\alpha,\alpha}^{2,1} & g_{\alpha,\alpha}^{2,2}
\end{array} \right] \right.\\
&-& \left. d \left[ \begin{array}{cc}
\Sigma_g^{2,2} & \Sigma_f \\
\Sigma_f & \Sigma_g^{1,1} 
\end{array} \right] \right\}
\nonb
,
\end{eqnarray}
where $d=g_{\alpha,\alpha}^{1,1} g_{\alpha,\alpha}^{2,2}
-\left(f_{\alpha,\alpha}^{1,2}\right)^2$ and
${\cal D} = 1 + 2 f_{\alpha,\alpha} \Sigma_f
-g_{\alpha,\alpha}^{1,1} \Sigma_g^{1,1}
-g_{\alpha,\alpha}^{2,2} \Sigma_g^{2,2}
+d \Sigma_d
$, with $\Sigma_d=\Sigma_g^{1,1} \Sigma_g^{2,2}
-\left( \Sigma_f \right)^2$.
The self-energies due to disorder take the form
\be
\left[\begin{array}{cc}
\Sigma_g^{1,1} & \Sigma_f^{1,2}\\
\Sigma_f^{1,2} & \Sigma_g^{2,2} \end{array} \right]
= \frac{n_\alpha u_\alpha^2 a_0^2}{(2\pi)^2}
\int d {\bf k} \left[ \begin{array}{cc}
g_{\alpha,\alpha}^{1,1}({\bf k}) & f_{\alpha,\alpha}^{1,2}({\bf k})\\
f_{\alpha,\alpha}^{1,2}({\bf k}) & g_{\alpha,\alpha}^{2,2}({\bf k})
\end{array} \right]
,
\ee
where $n_\alpha$ is the concentration of impurities and $u_\alpha$
is the scattering potential. We suppose that $u_\alpha$
is independent on wave vector.

The fully dressed Green's function
of the connected trilayer without vertex corrections
is obtained through
\be
\label{eq:Dyson1}
\HG_{\alpha,\alpha}=
\left[ \HI - 
 \Hg^{(d)}_{\alpha,\alpha} \Ht_{\alpha,a}
\Hg^{(d)}_{a,a}
\Ht_{a,\alpha}
- \Hg^{(d)}_{\alpha,\alpha} \Ht_{\alpha,b}
\Hg^{(d)}_{b,b}
\Ht_{b,\alpha}\right]^{-1}
\Hg^{(d)}_{\alpha,\alpha}
,
\ee
where $\HG_{\alpha,\alpha}$ stands for $\HG_{\alpha,\alpha}({\bf k},\omega)$,
and $\Hg^{(d)}_{i,i}$ stands for $\Hg^{(d)}_{i,i}({\bf k},\omega)$,
with $i=\alpha,a,b$.

\subsection{Vertex corrections}
\label{app:dis-vertex}
Lowest order vertex corrections
correspond to processes in which a quasiparticle
of the superconducting layer makes an excursion in one of the ferromagnetic
layers in between two scatterings on a given impurity in the
superconducting layer. 
The fully dressed Green's function is obtained through
\begin{eqnarray}
\label{eq:vertex}
&& \HG_{\alpha,\alpha} =\Hg^{(d)}_{\alpha,\alpha}\\\nonb
&+& \Hg^{(d)}_{\alpha,\alpha}\Ht_{\alpha,a}
\Hg^{(d)}_{a,a}
\Ht_{a,\alpha} \HG_{\alpha,\alpha}
+ \Hg^{(d)}_{\alpha,\alpha} \Ht_{\alpha,b}
\Hg^{(d)}_{b,b}
\Ht_{b,\alpha} \HG_{\alpha,\alpha}\\\nonb
&+&  n_\alpha a_0^2 \Hg^{(d)}_{\alpha,\alpha}
\int \frac{ d {\bf k}'}{(2\pi)^2}
\overline{\Hu_\alpha \Hg'^{(d)}_{\alpha,\alpha}
\Ht_{\alpha,a} \Hg'^{(d)}_{a,a}
\Ht_{a,\alpha} \Hg'^{(d)}_{\alpha,\alpha}
\Hu_\alpha} \HG_{\alpha,\alpha}\\\nonb
&+&  n_\alpha a_0^2 \Hg^{(d)}_{\alpha,\alpha}
\int \frac{ d {\bf k}'}{(2\pi)^2}
\overline{\Hu_\alpha \Hg'^{(d)}_{\alpha,\alpha}
\Ht_{\alpha,b} \Hg'^{(d)}_{b,b}
\Ht_{b,\alpha} \Hg'^{(d)}_{\alpha,\alpha}
\Hu_\alpha} \HG_{\alpha,\alpha}
,
\end{eqnarray}
where we used the same notation as for 
Eq.~(\ref{eq:Dyson1}) and
$g'^{(d)}_{i,i}=g^{(d)}_{i,i}({\bf k}',\omega)$,
$\Hu_\alpha=\Hu_\alpha({\bf k}-{\bf k}')$.

\section{Algorithm for the 1D model}
\label{app:algo}
In this Appendix we detail the algorithm by which we calculate the
energies of the Andreev bound states of the 1D model\cite{Vecino}.
In the superconductor the spectral representation of the local propagator
is obtained by summing Eq.~(\ref{eq:galpha11}) over all energy levels
of the 1D chain with open boundary conditions:
\begin{eqnarray}
&&g_{\alpha,\beta}^{1,1}(\omega)
=\frac{2 a_0}{L_S+a_0}
\sum_{n=1}^{L_S/a_0} 
\sin{\left(n \pi \frac{x_\alpha+a_0}{L_S+a_0}\right)}\\\nonb
&\times& \sin{\left(n \pi \frac{x_\beta+a_0}{L_S+a_0}\right)}
\left[ \frac{u_n^2}{\omega-E_n-i\eta}
+\frac{v_n^2}{\omega+E_n-i\eta} \right]
,
\end{eqnarray}
where $\alpha$ and $\beta$ correspond to two sites in the 1D chain
at positions $x_\alpha$ and $x_\beta$.
Similar expressions are obtained for the ``22'' and ``12'' components.
In the ferromagnet the energy levels are given by
$\epsilon_n^{(\sigma)}=\epsilon_n(L_F)-\sigma h_{\rm ex}$, where
$\epsilon_n(L)$ is given by Eq.~(\ref{eq:epsilon-n}).
The local Green's function at the extremity of the
1D ferromagnet is given by
\be
g_{a,a}^{(\sigma)}(\omega)
=\frac{2 a_0}{L_{F}+a_0} \sum_{n=1}^{L_{F}}
\sin^2{\left(\frac{n \pi a_0}{L_{F}+a_0}\right)}
\frac{1}{\omega-\epsilon_n^{(\sigma)}-i\eta}
.
\ee

We denote by $\alpha$ a site in the superconductor,
chosen far from the boundaries. At site $\alpha$ is
connected the extremity ``a'' of the ferromagnetic
chain. We note $t=t_{a,\alpha}=t_{\alpha,a}$.
The fully dressed Green's function 
$G_{\alpha,\alpha}^{1,1}$
of spin-up electrons
at site $\alpha$ is deduced from the
Dyson equations\cite{Melin-FSF,Jirari,Vecino}.
The spectral representation of $G_{\alpha,\alpha}^{1,1}$
is obtained
by evaluating numerically the position of the individual energy
levels $\omega_n$ and their spectral weights $R_n$: $G_{\alpha,\alpha}^{1,1}
=\sum_n R_n^{1,1} / (\omega-\omega_n-i\eta)$.
In the limit $\eta \rightarrow 0$ the LDOS is
given by 
\be
\rho_{\alpha,\alpha}^{1,1}(\omega) =
\frac{1}{\pi} \mbox{Im}\left[ G_{\alpha,\alpha}^{1,1} (\omega)\right] =
\sum_n R_n^{1,1}
\delta(\omega-\omega_n)
.
\ee
The energy levels and spectral weights can be obtained
without approximation to an
arbitrary precision by using a dichotomy algorithm.


\begin{thebibliography}{99}

\bibitem{Tinkham} M. Tinkham, {\it Introduction to
superconductivity} (Mc Graw-Hill, 1996).

\bibitem{Tedrow} P. Tedrow and R. Meservey,
Phys. Rev. Lett. {\bf 26}, 192 (1971);
Phys. Rev. B {\bf 7}, 318 (1973);
R. Meservey and P. Tedrow, Phys. Rep. {\bf 238},
173 (1994).
 
\bibitem{Soulen} R. J. Soulen {\sl et al.},
Science {\bf 282}, 85 (1998).

\bibitem{Upadhyay} S. K. Upadhyay, A. Palanisami, R. N. Louie,
and R. A. Buhrman, Phys. Rev. Lett. {\bf 81}, 3247 (1998).

\bibitem{deJong-Beenakker} M.J.M. de Jong and C. W.
Beenakker, Phys. Rev. Lett. {\bf 74},
1657 (1995).

\bibitem{Falko} V.I. Fal'ko, C.J. Lambert and A.F. Volkov
Pis'ma Zh. \'Eksp. Teor. Fiz. {\bf 69}, 497 (1999)
[JETP Letters {\bf 69}, 532 (1999)].

\bibitem{Jedema} F.J. Jedema,
B.J. van Wees, B.H. Hoving, A.T. Filip and T.M. Klapwijk,
Phys. Rev. B {\bf 60}, 16549 (1999).

\bibitem{Belzig} W. Belzig, A. Brataas, Yu. V. Nazarov
and G.E.W. Bauer, Phys. Rev. B {\bf 62}, 9726 (2000).

\bibitem{Melin-Europhys} R. M\'elin, Europhys. Lett.
{\bf 51}, 202 (2000).

\bibitem{Fulde-Ferrel} P. Fulde and A. Ferrel,
Phys. Rev. {\bf 135}, A550 (1964).

\bibitem{Larkin} A. Larkin and Y. Ovchinnikov,
Zh. Eksp. Teor. Fiz. {\bf 47}, 1136 (1964)
[Sov. Phys. JETP {\bf 20}, 762 (1965)].

\bibitem{Clogston} M.A. Clogston,
Phys. Rev. Lett. {\bf 9}, 266 (1962).

\bibitem{Demler} E.A. Demler,
G.B. Arnold and M.R. Beasley, Phys. Rev. B
{\bf 55}, 15174 (1997).

\bibitem{Buzdin1} A.I. Buzdin, L.N. Bulaevskii,
and S.V. Panyukov, Pis'ma Zh. Eksp. Teor.
Fiz. {\bf 35}, 147 (1982)
[JETP Lett. {\bf 35},
178 (1982)];
A. Buzdin, B. Bujicic, and M. Yu. Kupriyanov,
Zh. Eksp. Teor. Fiz.
{\bf 101}, 231 (1992)
[Sov. Phys. JETP {\bf 74}, 124 (1992)].

\bibitem{Buzdin2}
A.I. Buzdin and M. Yu. Kupriyanov,
Pis'ma Zh. Eksp. Teor. Fiz. {\bf 52}, 1089 (1990)
[JETP Lett. {\bf 52}, 487 (1990)];
A.I. Buzdin, M. Yu.
Kupriyanov and B. Vujicic,
Physica C {\bf 185 - 189}, 2025 (1991).

\bibitem{Radovic} Z. Radovi\'c, M. Ledvij,
L. Dobrosavljevi\'c, A.I. Buzdin,and J. R. Clem, Phys. Rev. B
{\bf 44}, 759 (1991).

\bibitem{exp1} J.S. Jiang, D. Davidovi\'c,
D.H. Reich, and C.L. Chien,
Phys. Rev. Lett. {\bf 74}, 314 (1995);
J.S. Jiang, D. Davidovi\'c,
D.H. Reich, and C.L. Chien, Phys. Rev. B {\bf 54},
6119 (1996);
C.L. Chien, J.S. Jiang,
J.Q. Xiao, D. Davidovi\'c, and
D.H. Reich, J. Appl. Phys. {\bf 81},
5358 (1997).

\bibitem{exp2} L.V. Mercaldo, C. Attanasio,
C. Coccorese, L. Maritato,
S.L. Prischepa, and M. Salvato,
Phys. Rev. B {\bf 53}, 14 040 (1996).

\bibitem{exp3} Th. M\"uhge, N.N. Garif'yanov,
Yu. V. Goryunov, G.G. Khaliullin, L.R. Tagirov,
K. Westerholt, I.A. Garifullin, and
H. Zabel, Phys. Rev. Lett. {\bf 77}, 1857 (1996).

Th. Muhge, K. Westerholt, H. Zabel, N.N.
Garif'yanov, Yu. V. Goryunov, I.A. Garifullin, and
G.G. Khaliullin, Phys. Rev. B {\bf 55}, 8945 (1997).

\bibitem{transpa} J. Aartz, J.M.E. Geers, E. Br\"uck,
A.A. Golubov and R. Coehoorn,
Phys. Rev. B {\bf 56}, 2779 (1997);
L. Lazar, K. Westerholt, H. Zabel, L.R. Tagirov,
Yu. V. Goryunov, N.N. Garif'yanov
and I.A. Garifullin, {\it ibid.}
{\bf 61}, 3711 (2000);
M. Sch\"ock, C. S\"urgers and H.v. L\"ohneysen,
Eur. Phys. J. B {\bf 14}, 1 (2000).

\bibitem{Zareyan} M. Zareyan, W. Belzig and Yu. V. Nazarov,
Phys. Rev. Lett. {\bf 86}, 308 (2001);
Phys. Rev. B {\bf 65}, 184505 (2002).

\bibitem{Ryazanov} V.V. Ryazanov, V.A. Oboznov,
A. Yu. Rusanov, A.V. Veretennikov,
A.A. Golubov, J. Aarts, Phys. Rev. Lett.
{\bf 86}, 2427 (2001).

\bibitem{Kontos} T. Kontos, M. Aprili,
J. Lesueur, and X. Grison,
Phys. Rev. Lett. {\bf 86}, 304 (2001).

\bibitem{Gandit} W. Guichard, M. Aprili, O. Bourgeois, T. Kontos,
J. Lesueur and P. Gandit, Phys. Rev. Lett. {\bf 90}, 167001 (2003).

\bibitem{Sellier} H. Sellier, C. Baraduc, F. Lefloch
and R. Calemczuk, Phys. Rev. B {\bf 68}, 054531 (2003).

\bibitem{deGennes-bound} P.G. de Gennes and
D. Saint-James, Phys. Letters {\bf 4}, 151 (1963);
D. Saint-James, Journal de Physique {\bf 25}, 899 (1964).

\bibitem{Andreev} A.F. Andreev, Sov. Phys. JETP {\bf 19},
1228 (1964).

\bibitem{Vecino} E. Vecino, A. Martin-Rodero and
A. Levy Yeyati, Phys. Rev. B {\bf 64}, 184502 (2001).

\bibitem{Cserti} J. Cserti, J. Koltai and C.J. Lambert,
cond-mat/0306523.

\bibitem{Valls} K. Halterman and O. T. Valls,
Phys. Rev. B {\bf 66}, 224516 (2002).

\bibitem{deGennes-FSF} P.G. de Gennes, Phys.
Letters {\bf 23}, 10 (1966).

\bibitem{Deutscher-Meunier} G. Deutscher and F. Meunier,
Phys. Rev. Lett. {\bf 22}, 395 (1969).

\bibitem{Hauser} J.J. Hauser, Phys. Rev. Lett.
{\bf 23}, 374 (1969).

\bibitem{Gu} J. Y. Gu, C.-Y. You, J. S. Jiang, J. Pearson,
Ya. B. Bazaliy, and S. D. Bader,
Phys. Rev. Lett. 89, 267001 (2002).

\bibitem{Volkov} F.S. Bergeret, A.F. Volkov and K.B. Efetov,
cond-mat/0307468.

\bibitem{Jirari}
H. Jirari, R. M\'elin and N. Stefanakis, Eur. Phys. J. B
{\bf 31}, 125 (2003).

\bibitem{Melin-FSF} R. M\'elin, J. Phys.: Condens. Matter
{\bf 13}, 6445 (2001);
V. Apinyan and R. M\'elin,
Eur. Phys. J. B {\bf 25}, 373 (2002);

\bibitem{Baladie} I. Baladi\'e and A. Buzdin,
Phys. Rev. B {\bf 67}, 014523 (2003);
I. Baladi\'e, A. Buzdin,
N. Ryzhanova, and A. Vedyayev,
Phys. Rev. B {\bf 63}, 054518 (2001);
A. Buzdin, A.V. Vedyayev, and N. Ryzhanova,
Europhys. Lett. {\bf 48}, 686 (1999).

\bibitem{Buzdin-Daumens} A. Buzdin and M. Daumens, 
Europhys. Lett. {\bf 64}, 510 (2003).

\bibitem{Melin-Feinberg-FSF} R. M\'elin and D. Feinberg,
cond-mat/0308007, Europhys. Lett. in press.

\bibitem{Deutscher-Feinberg} G. Deutcher and D. Feinberg, Appl. Phys. Lett. 
{\bf 76}, 487 (2000).

\bibitem{Falci} G. Falci, D. Feinberg and F.W.J. Hekking,
Europhys. Lett. {\bf 54}, 255 (20001).

\bibitem{Melin-Feinberg-tr} R. M\'elin and D. Feinberg,
Eur. Phys. J. B {\bf 26}, 101 (2002).

\bibitem{Melin-Peysson} R. M\'elin and S. Peysson,
cond-mat/0302236, Phys. Rev. B in press.

\bibitem{Lambert} 
C.J. Lambert, J. Phys. Condens. Matter
{\bf 3} 6579 (1991); 
C.J. Lambert and R. Raimondi, J. Phys. Condens. Matter
{\bf 10} 901 (1998).

\bibitem{Andreev-Buzdin} A.V. Andreev, A.I. Buzdin and R.M.
Osgood III, Phys. Rev. B {\bf 43}, 10124 (1991).

\bibitem{Houzet} M. Houzet and A. Buzdin,
Europhys. Lett. {\bf 58}, 596 (2002).

\bibitem{Fominov} Ya. V. Fominov, N.M. Chtchelkatchev
and A.A. Golubov, Ois'ma Zh. Eksp. Teor. Fiz.
{\bf 74}, 101 (2001) [JETP Lett. {\bf 74},
96 (2001)]; Phys. Rev. B {\bf 66},
014507 (2002).

\bibitem{Khusainov} Yu. N. Proshin and M.G. Khusainov, Zh. Eksp. Teor. Fiz.
{\bf 113}, 1708 (1998), {\bf 116}, 1887 (1999) [JETP
{\bf 86}, 930 (1998); {\bf 89}, 1021 (1999)];
M.G. Khusainov and Yu. N. Proshin, Phys. Rev. B {\bf 56},
R14283 (1997); {\bf 62}, 6832 (2000).

\bibitem{Bergeret-2001} F.S. Bergeret, A.F. Volkov and K.B. Efetov,
Phys. Rev. Lett. {\bf 18}, 4096 (2001).

\bibitem{Kadi} A. Kadigrobov, R.I. Shekhter and M. Jonson,
Europhys. Lett. {bf 54}, 394 (2001).

\bibitem{Chte} N.M. Chtchelkatchev and I. Burmistrov,
Phys. Rev. B {\bf 68}, 140501 (2003).

\bibitem{Abrikosov} A.A. Abrikosov, L.P. Gorkov and I.E.
Dzyaloshinski, {\sl Methods of quantum field theory in
statistical physics}, Dover Publications, Inc, New York (1963).

\bibitem{Belzig-quasi} W. Belzig, C. Bruder and G. Sch\"on,
Phys. Rev. B {\bf 54}, 9443 (1996).

\bibitem{note} In Ref.~\onlinecite{Vecino} the Green's functions
of the ferromagnet are real numbers. The spin-up and spin-down
density of states are thus vanishingly small whereas the opposite
is expected for a metallic ferromagnet.

\bibitem{BTK} G.E. Blonder, M. Tinkham and T.M. Klapwijk,
Phys. Rev. B {\bf 25}, 4515 (1982).
\end{thebibliography}
\end{document}